\documentclass[aps,prm,twocolumn,longbibliography,superscriptaddress,nofootinbib]{revtex4-1}

\usepackage[colorlinks=true,citecolor=blue,linkcolor=blue]{hyperref}
\usepackage{graphicx}
\usepackage{xcolor}
\usepackage{bm,mathbbol}
\usepackage{units}

\renewcommand{\vec}[1]{\ensuremath{\mathbf{#1}}}

\begin{document}

\title{Non-trivial topological valence bands of common diamond and zinc-blende semiconductors}

\author{Tom\'{a}\v{s} Rauch} \affiliation{Friedrich-Schiller-University Jena, 07743 Jena, Germany}

\author{Victor A. Rogalev}
\affiliation{\mbox{Physikalisches Institut and W\"urzburg-Dresden Cluster of Excellence  ct.qmat, Universit\"at W\"urzburg, 97074 W\"urzburg, Germany}}

\author{Maximilian Bauernfeind}
\affiliation{\mbox{Physikalisches Institut and W\"urzburg-Dresden Cluster of Excellence  ct.qmat, Universit\"at W\"urzburg, 97074 W\"urzburg, Germany}}

\author{Julian Maklar}
\affiliation{\mbox{Physikalisches Institut and W\"urzburg-Dresden Cluster of Excellence  ct.qmat, Universit\"at W\"urzburg, 97074 W\"urzburg, Germany}}

\author{Felix Reis}
\affiliation{\mbox{Physikalisches Institut and W\"urzburg-Dresden Cluster of Excellence  ct.qmat, Universit\"at W\"urzburg, 97074 W\"urzburg, Germany}}

\author{Florian Adler}
\affiliation{\mbox{Physikalisches Institut and W\"urzburg-Dresden Cluster of Excellence  ct.qmat, Universit\"at W\"urzburg, 97074 W\"urzburg, Germany}}

\author{Simon Moser}
\affiliation{\mbox{Physikalisches Institut and W\"urzburg-Dresden Cluster of Excellence  ct.qmat, Universit\"at W\"urzburg, 97074 W\"urzburg, Germany}}

\author{Johannes Weis}
\affiliation{\mbox{Physikalisches Institut and W\"urzburg-Dresden Cluster of Excellence  ct.qmat, Universit\"at W\"urzburg, 97074 W\"urzburg, Germany}}

\author{Tien-Lin Lee}
\affiliation{Diamond House, Harwell Science and Innovation Campus, Didcot, Oxfordshire OX11 0DE, United Kingdom}

\author{Pardeep K. Thakur}
\affiliation{Diamond House, Harwell Science and Innovation Campus, Didcot, Oxfordshire OX11 0DE, United Kingdom}

\author{J\"{o}rg Sch{\"a}fer}
\affiliation{\mbox{Physikalisches Institut and W\"urzburg-Dresden Cluster of Excellence  ct.qmat, Universit\"at W\"urzburg, 97074 W\"urzburg, Germany}}

\author{Ralph Claessen}
\affiliation{\mbox{Physikalisches Institut and W\"urzburg-Dresden Cluster of Excellence  ct.qmat, Universit\"at W\"urzburg, 97074 W\"urzburg, Germany}}

\author{J\"{u}rgen Henk} \affiliation{Institute of Physics, Martin Luther University Halle-Wittenberg, Halle (Saale), Germany}

\author{Ingrid Mertig} \affiliation{Institute of Physics, Martin Luther University Halle-Wittenberg, Halle (Saale), Germany}
\affiliation{Max Planck Institute for Microstructure Physics, Halle (Saale), Germany}

\date{\today}

\begin{abstract}
The diamond and zinc-blende semiconductors are well-known and have been widely studied for decades. Yet, their electronic structure still surprises with unexpected topological properties of the valence bands. In this joint theoretical and experimental investigation we demonstrate for the benchmark compounds InSb and GaAs that the electronic structure features topological surface states below the Fermi energy. Our parity analysis shows that the spin-orbit split-off band near the valence band maximum exhibits a strong topologically non-trivial behavior characterized by the $\mathcal{Z}_2$ invariants $(1;000)$. The non-trivial character emerges instantaneously with non-zero spin-orbit coupling, in contrast to the conventional topological phase transition mechanism. \textit{Ab initio}-based tight-binding calculations resolve topological surface states in the occupied electronic structure of InSb and GaAs, further confirmed experimentally by soft X-ray angle-resolved photoemission from both materials. Our findings are valid for all other materials whose valence bands are adiabatically linked to those of InSb, i.e., many diamond and zinc-blende semiconductors, as well as other related materials, such as half-Heusler compounds.
\end{abstract}

\maketitle

\section{Introduction}
\label{sec:intro}
The prediction of two-dimensional time-reversal invariant topological insulators (TI)~\cite{Kane2005,Kane2005a,Bernevig2006} triggered the ongoing search for topologically non-trivial materials. The current classification of condensed matter includes a huge variety of different topological classes, such as three-dimensional (3D) TIs~\cite{Moore2007,Fu2007,Roy2010}, topological crystalline insulators in two~\cite{Wrasse2014,Liu2015} and three~\cite{Fu2011,Hsieh2012} dimensions, Weyl~\cite{Murakami2007,Wan2011,Yan2017}, Dirac~\cite{Young2012,Yang2014a,Yang2015} and nodal-line semimetals~\cite{Burkov2011}, and many more.

According to recent developments, non-trivial topology is not restricted to energies close to the Fermi energy; it can be found all over the band structure and seems to be more common than previously believed~\cite{Bradlyn2017}. This observation opens a challenge for both theory and experiment to identify topologically non-trivial states in many well-known compounds, including those with a trivial fundamental band gap. One example is the recently discovered non-trivial topology  in double-layers of traditional semiconductors~\cite{Lucking2018}.

In this paper we report on a joint theoretical and experimental investigation of diamond and zinc-blende semiconductors whose notorious band structure hosts a \textit{trivial fundamental band gap}. For this well understood class of materials we show that non-trivial topology is hidden just below the topmost valence band; more precisely, we focus on the spin-orbit induced gap (SO-gap) among the occupied $p$-bands.
InSb and GaAs serve as exemplary materials to study due to the extraordinarily large SO-gap of the former and the familiarity of the latter.

We utilize an \textit{ab initio}-based tight-binding (TB) model to calculate the $\mathcal{Z}_2$ topological invariants and the surface electronic structure of a semi-infinite InSb crystal to reveal the associated topological surface states (TSSs). The latter is computed also for GaAs using a proven TB parameterization~\cite{Vogl1983}. On top of this, soft X-ray angle-resolved photoelectron spectroscopy (SX-ARPES) for the (001) surface of InSb and GaAs reveals the TSSs among the projected valence bands.

Our results hold also for diamond and zinc-blende semiconductors with a non-trivial fundamental band gap, such as strained $\alpha$-Sn and HgTe~\cite{Fu2007a}, for which a another TSS was recently revealed in the valence band in addition to the established TSS at the Fermi level~\cite{Rogalev2017}.

\section{Theoretical aspects}
\label{sec:theory}
Diamond and zinc-blende materials crystallize in a face-centered cubic lattice with a two-atomic basis. The two basis atoms are identical in the case of a diamond lattice (space group 227, Wyckoff position 8a: $(1/8,1/8,1/8)$); typical materials are Si and Ge. However, the basis atoms are different for the zinc-blende structure (space group 216, Wyckoff positions 4a: $(0,0,0)$ and 4c: $(1/4,1/4,1/4)$), with the  important consequence that inversion symmetry is broken; typical compounds are GaAs, CdTe, and InSb.

\begin{figure}
\centering
\includegraphics[width=1.0\columnwidth]{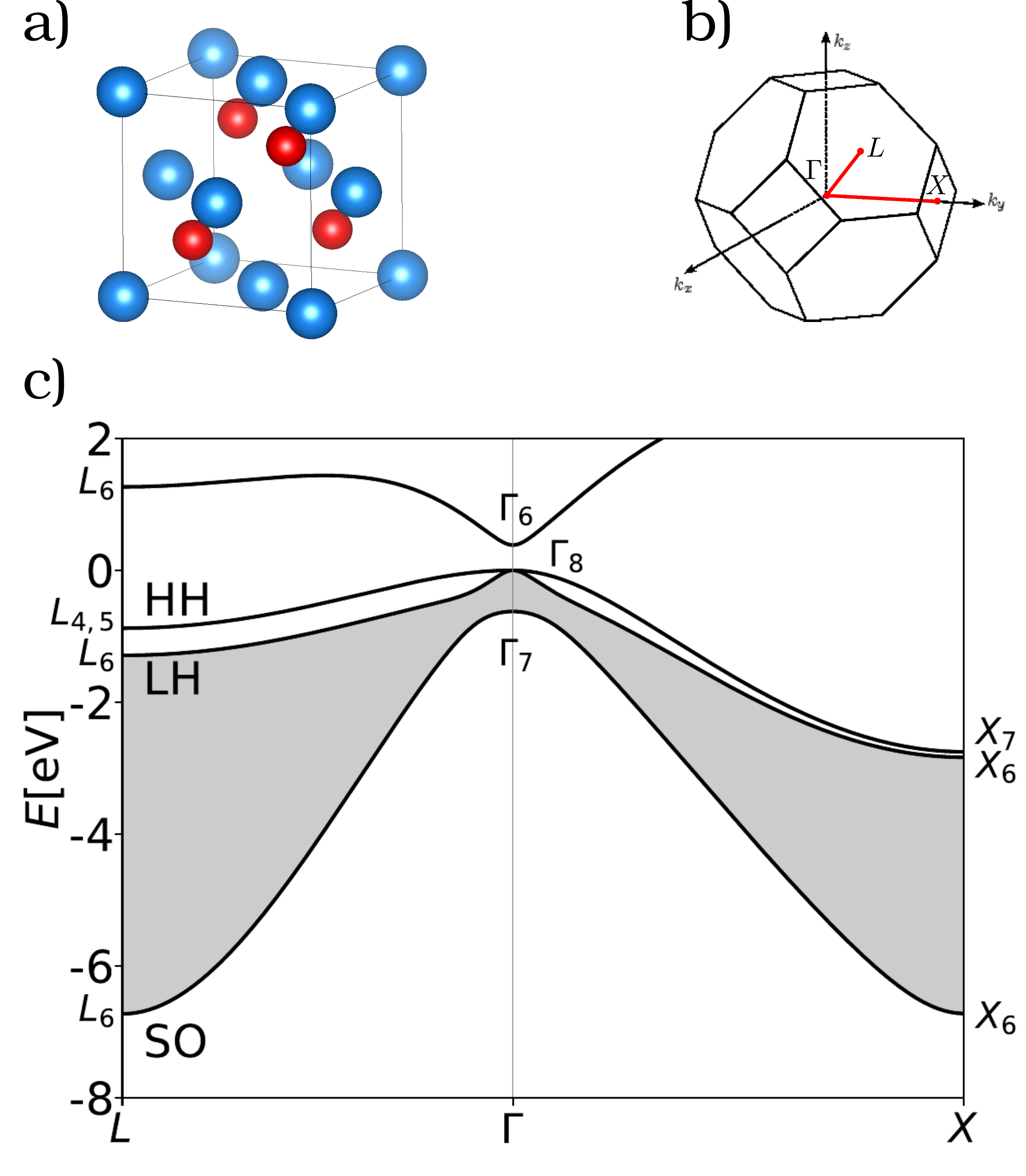}
\caption{InSb in zinc-blende lattice. (a) Crystal structure (drawn with VESTA~\cite{Momma2011}). (b) Brillouin zone with TRIMs $\Gamma$, $L$, and $X$. (c) Band structure along the path shown red in (b). HH, LH, and SO denote the heavy-hole, light-hole, and spin-orbit split band, respectively. The shaded area marks the global topologically nontrivial SO-gap.}
\label{fig:bands_bulk}
\end{figure}

The electronic structures of these solids share a number of features, for example a topologically trivial fundamental band gap~\footnote{Exceptions are $\alpha$-Sn and HgTe, which we explicitly exclude from the discussions unless stated otherwise.}. The two lowest conduction bands are derived from $s$ orbitals, while the six valence bands are $p$-derived. Without spin-orbit coupling (SOC) the latter are degenerate at the center $\Gamma$ of the Brillouin zone (BZ). If SOC is included this six-fold degeneracy is split into a four-fold degenerate and a two-fold degenerate level (`spin-orbit split-off band', SO for short), with the latter (often) lower in energy. The four-fold degenerate band splits further into the  light- and heavy-hole bands away from $\Gamma$ (abbreviated LH and HH, respectively; Fig.~\ref{fig:bands_bulk}). These features of many diamond and zinc-blende materials are well-known and have been studied for the bulk and for surfaces  \cite{Yu2010fundamentals,Liu2015HgTe,Scholz2018,Morgenstern2012}.

We focus on the energy range between the LH and the SO band (shaded area in Fig.~\ref{fig:bands_bulk}c). This SO-gap extends throughout the entire BZ, which allows to characterize the topological properties of the bands below the LH and HH pair. For $\alpha$-Sn~\cite{Rogalev2017} and Hg$_x$Cd$_{1-x}$Te~\cite{Rauch2017} it has been shown that the LH and the HH bands as well as those below the SO-gap are topologically non-trivial. Here we show that also semiconductors with a trivial fundamental band gap can host a topologically non-trivial SO-gap. A first hint for this can be found in Ref.~\onlinecite{Ohtsubo2012} which reports on a study of  the surface electronic structure of Ge. The authors mention and demonstrate (cf.\ Fig.~2a of that paper) that the LH- and SO-character of the bands near $\Gamma$ become inverted for large $\vec{k}$, suggesting a band inversion typical for topological insulators.  Furthermore, a surface resonance state of (001)-terminated Ge located in the projected SO-gap has been reported in Ref.~\onlinecite{Seo2014} (cf.\ Fig.~4a of that paper). This feature might be a TSS and we consider it another clue for the non-trivial topology of the SO-gap.

We want to emphasize that the band inversion in the SO-gap of zinc-blende and diamond semiconductors is instantaneous in the sense that it exists as soon as the SOC strength is turned non-zero (in a theoretical treatment in which the SOC strength can be tuned). This stands in contrast to the common belief that a band inversion must occur via tuning a continuous parameter (e.\,g., SOC strength or lattice constant); the topological phase transition is then accompanied by closing and reopening of the gap, as is found in, e.\,g., Bi$_2$Te$_3$. These findings are in agreement with the recent concept of topological quantum chemistry~\cite{Bradlyn2017,Vergniory2018}, where the two possible scenarios for topologically non-trivial band gaps occur.

We take InSb as a first exemplary material; its large SO-gap of about  $\unit[0.8]{eV}$ at $\Gamma$ facilitates the theoretical and experimental identification of TSSs within the surface-projected bands. Choosing GaAs as a second well-known example, we demonstrate that the TSSs can be observed also in the much narrower SO-gap of this material.

\subsection{Electronic structure and topological characterization of InSb}
For the  calculations of the electronic structure of InSb we employ an $sp^{3}$ TB model with first-  and second-nearest neighbor coupling. The on-site energies, hopping amplitudes, and the SOC strengths in the Slater-Koster parameterization (Table~\ref{tab:sk-pars}) are optimized by adjusting the TB bands  to those calculated within density-functional theory using highly accurate hybrid functionals~\cite{Kim2009}.

\begin{table}
 \caption{Slater-Koster parameters of InSb used for the TB model. All numbers in eV.} \label{tab:sk-pars}
 \centering
 \begin{tabular}{l|c|c}
 	\hline \hline
 	On-site energies & $E^s_{\mathrm{In}}$ & -6.55 \\
 	& $E^p_{\mathrm{In}}$ &  0.48 \\
 	& $E^s_{\mathrm{Sb}}$ & -8.84 \\
 	& $E^p_{\mathrm{Sb}}$ &  1.48 \\
	\hline
 	SO strength & $\lambda_{\mathrm{In}}$ &  0.14 \\
 	& $\lambda_{\mathrm{Sb}}$ &  0.26 \\
 	\hline
 	1NN & $(ss\sigma)_{\mathrm{In-Sb}}$ & -2.71 \\
 	& $(sp\sigma)_{\mathrm{In-Sb}}$ & -1.42 \\
 	& $(ps\sigma)_{\mathrm{In-Sb}}$ &  2.71 \\
 	& $(pp\sigma)_{\mathrm{In-Sb}}$ &  2.65 \\
 	& $(pp\pi)_{\mathrm{In-Sb}}$ & -0.64 \\
  	\hline
 	2NN & $(ss\sigma)_{\mathrm{In-In}}$ & -0.24 \\
 	& $(sp\sigma)_{\mathrm{In-In}}$ &  0.01 \\
 	& $(pp\sigma)_{\mathrm{In-In}}$ &  0.47 \\
 	& $(pp\pi)_{\mathrm{In-In}}$ & -0.10 \\
  	\hline
 	2NN & $(ss\sigma)_{\mathrm{Sb-Sb}}$ & -0.28 \\
 	& $(sp\sigma)_{\mathrm{Sb-Sb}}$ &  0.0 \\
 	& $(pp\sigma)_{\mathrm{Sb-Sb}}$ &  0.14 \\
 	& $(pp\pi)_{\mathrm{Sb-Sb}}$ & -0.10 \\
  	\hline  \hline
 \end{tabular}
\end{table}

In a first step we determine the joint $\mathcal{Z}_2$ invariants of all bands below the LH bands, that are the SO bands and the lower $s$-derived bands (bottom of the valence bands, not included in Fig.~\ref{fig:bands_bulk}), using two approaches.

The first approach relies on the fact that the occupied bands of InSb can be adiabatically linked to those of a material with diamond lattice (e.\,g., Ge). Since the gaps between the individual bands do not close during the transformation, both materials belong to the same topological class. The double group representations of the bands below the SO-gap of the zinc-blende and diamond lattices at the time-reversal invariant momenta (TRIMs) are connected as: $\left(L_6, L_6\right) \rightarrow \left(L^{-}_6, L^{+}_6\right)$, $\left(\Gamma_6, \Gamma_7\right) \rightarrow \left(\Gamma^{+}_6, \Gamma^{+}_7\right)$, $\left(X_6, X_6\right) \rightarrow \left(X_5\right)$~\cite{Chelikowsky1976,Dresselhaus2008}. Since the diamond structure is inversion symmetric (in contrast to InSb), multiplying the parities of the individual bands at the TRIMs provides the $\mathcal{Z}_2$ invariants~\cite{Fu2007a}.

The $X$ point of the BZ of diamond materials requires special treatment. There, only the double-group representation $X_5$ exists, which corresponds to a four-fold degenerate level. Consulting the Bilbao Crystallographic Server~\cite{Elcoro2017} one finds that the matrix of $X_5$ corresponding to the inversion has diagonal elements $\left(1,1,-1,-1\right)$. Since this is a feature of the diamond structure, the required parities can be obtained by imagining a distortion of the lattice while preserving the inversion symmetry. This lifts the four-fold degeneracy and the parities of the two remaining doubly degenerate levels are $1$ and $-1$, giving $-1$ for the product of the parities. This method was adopted from Ref.~\onlinecite{Fu2007a}. 

From the multiplicities and the parity products at the TRIMS (Table~\ref{tab:parities})  we deduce $\mathcal{Z}_2$ invariants of $(1;000)$~\cite{Fu2007a}, which implies that the group of bands below the LH band is topologically non-trivial and behaves as a strong TI\@.

\begin{table}
 \caption{TRIMs, multiplicity, double group representations and products of parities.} \label{tab:parities}
 \centering
 \begin{tabular}{cccc}
 	\hline
 	\hline
 	TRIM & multiplicity & representation & parity product \\
 	\hline
 	$\Gamma$ & 1 & $\Gamma^{+}_{6}$ $\Gamma^{+}_{7}$ & $+$ \\
 	X & 3 & $X_5$ & $-$ \\
 	L & 4 & $L^{-}_{6}$ $L^{+}_{6}$ & $-$ \\
 	\hline
  	\hline
 \end{tabular}
\end{table}

To confirm the above result (which is  restricted to inversion symmetric crystals) we calculate the $\mathcal{Z}_2$ invariants by the Wannier sheets (WSs) approach, which utilizes the eigenvalues of the position operator along a given direction projected onto the individual bands~\cite{Yu2011}. The WSs are obtained by calculating the Berry phases along a periodic direction in reciprocal space, with the coordinates along the other two axes fixed (that is a  Wilson loop). As shown in Ref.~\onlinecite{Taherinejad2014}, this method allows us to visualize the partner switching of the WSs and thus the $\mathcal{Z}_2$ invariants.

\begin{figure}
	\centering
	\includegraphics[width=0.9\columnwidth]{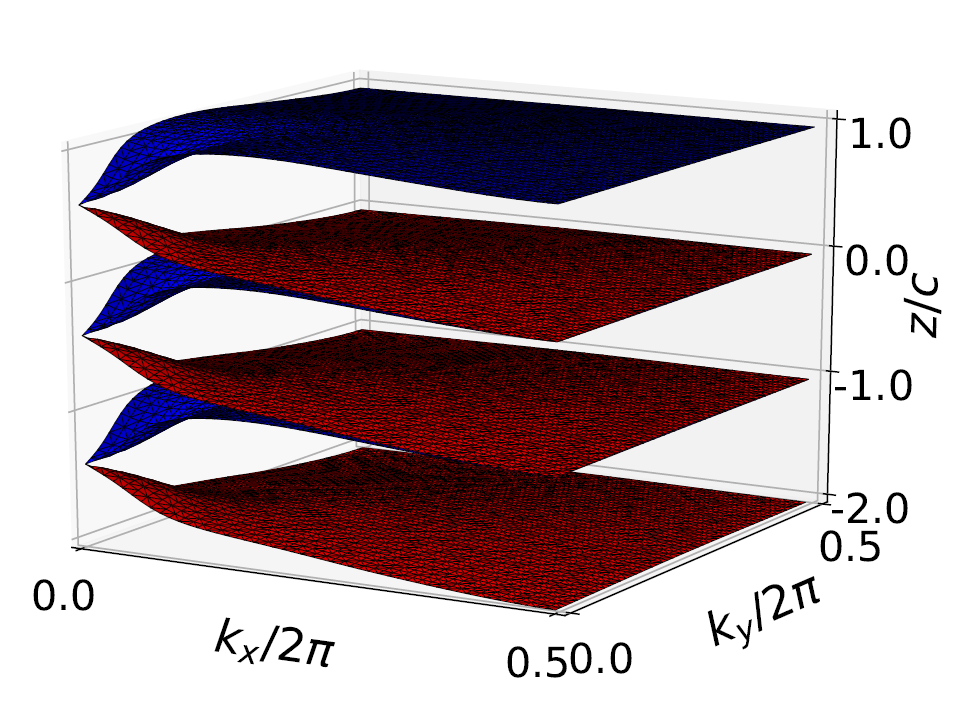}
	\includegraphics[width=0.9\columnwidth]{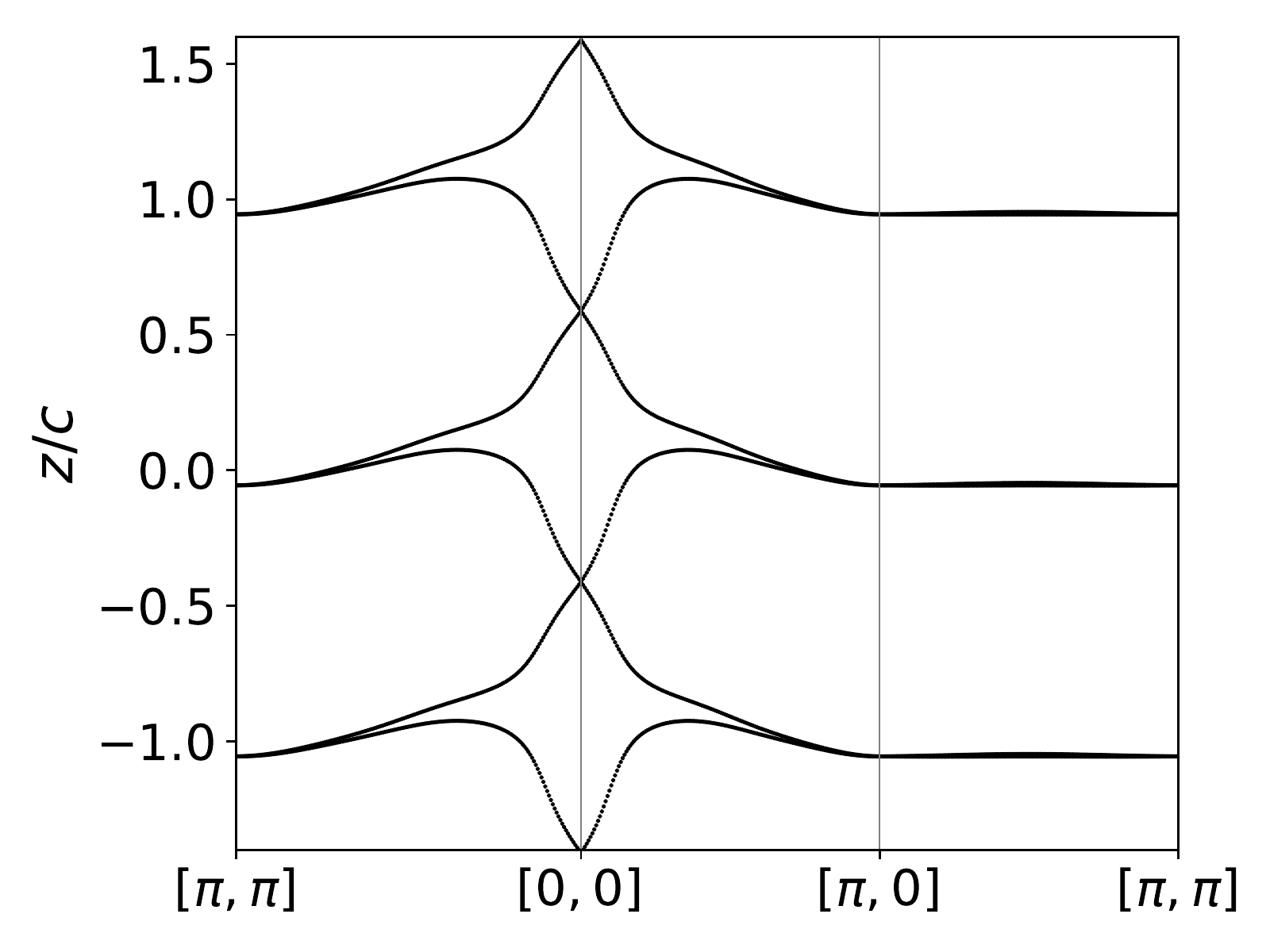}
	\caption{Periodic Wannier sheets (WSs) constructed from individuals bands of InSb. $k_{x}$ and $k_{y}$ are axes oriented along the reciprocal lattice vectors $\vec{G}_{1}$ and $\vec{G}_{2}$, respectively. The Wilson loops were calculated along $\vec{G}_{3}$.  The $z$  coordinate is along the [110] direction, that is  perpendicular to both $\vec{G}_{1}$ and $\vec{G}_{2}$, $c$ is the distance between two unit cells along this direction. Top: WSs in \nicefrac{1}{4} of the $\vec{G}_{1}\vec{G}_{2}$ plane with the TRIMs at the corners. Bottom: WSs along lines connecting the TRIMs X--$\Gamma$--L--X\@.}
	\label{fig:WSs}
\end{figure}

We calculate the WSs in a plane through $\Gamma$  which is spanned by $\vec{G}_{1} = 2\pi (-1,1,1)$ and $\vec{G}_{2}= 2\pi (1,-1,1)$ (Fig.~\ref{fig:WSs}); $\vec{G}_{3}= 2\pi (1,1,-1)$ marks the direction of the Wilson loop, setting the lattice constant $a=1$ for simplicity. Partner switching of the WSs shows up at $k_x = 0$ and $k_y = 0$, and there is no switching at $k_x = \pi$ and $k_y = \pi$. This means all three weak topological invariants are $0$ and the strong one equals $1$; hence, $\mathcal{Z}_2 = (1;000)$ which confirms our first result.

\subsection{Surface electronic structure of InSb and GaAs}
Having established the non-trivial topology of the SO bulk band, we aim at finding TSSs that have to connect the SO band with the LH or HH bands projected on \emph{any} surface. For this purpose we choose for our TB model a semi-infinite geometry with ideal (001) surface (uncovered and unreconstructed; recall that TSSs exist irrespectively of the surface details). The layer- and wavevector-resolved spectral density is computed by Green function renormalization~\cite{Henk1993,Bodicker1994}.

\begin{figure}
	\centering
	\includegraphics[width=\columnwidth]{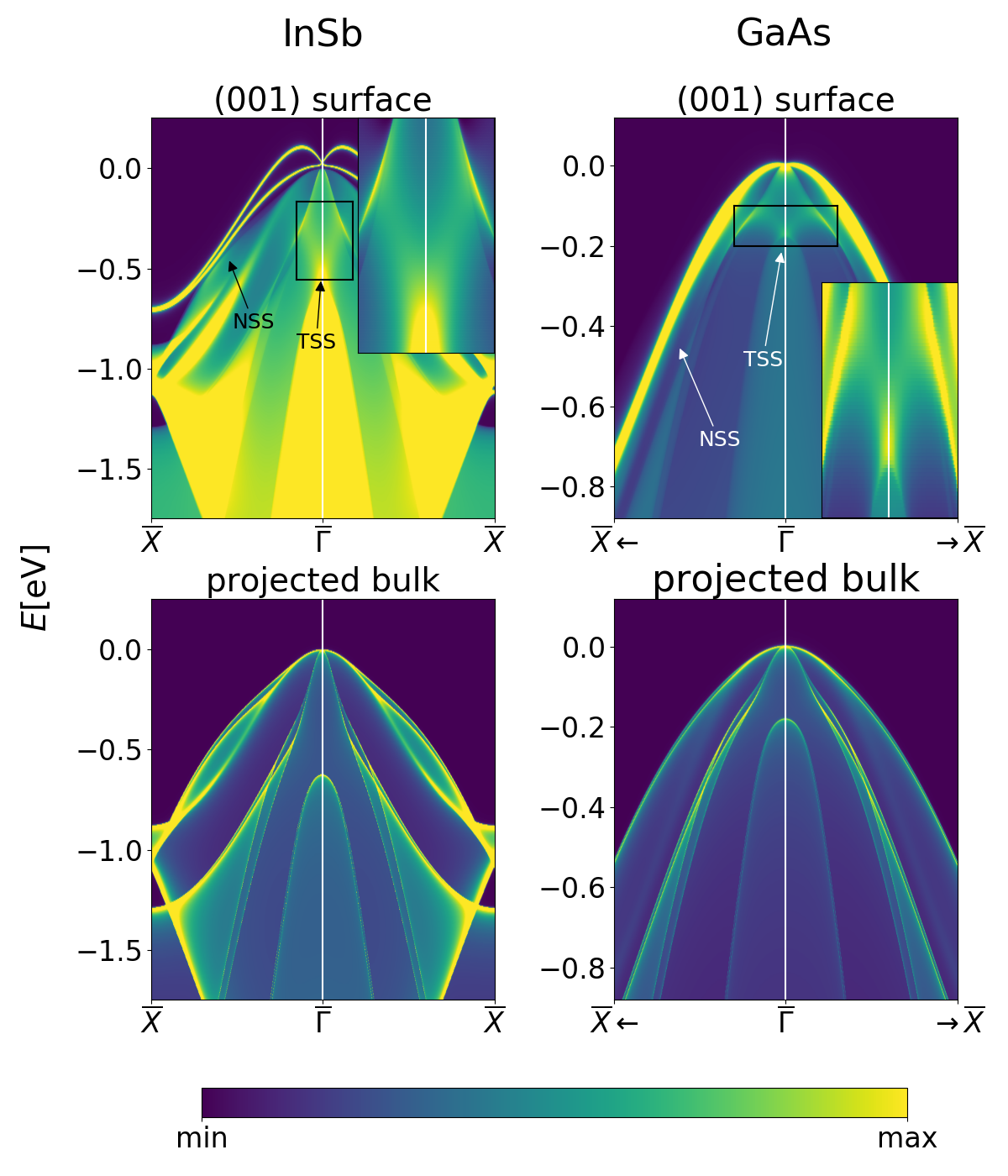}
	\caption{Theoretical electronic structure of ideal InSb (left) and GaAs (right) (001) surfaces. The spectral density of the topmost layer (top) and of a bulk layer (bottom) is depicted as color scale. Positions of the topological surface state (TSS) and a normal surface state (NSS) are highlighted by arrows.}
	\label{fig:surf}
\end{figure}

The LH and the HH bulk bands projected onto the surface plane cover the entire SO-gap (bottom in Fig.~\ref{fig:surf}). As a result, a TSS becomes strictly speaking a surface resonance which shows up as a broad feature in the spectral density, in contrast to a sharp feature for a true surface state. Although this hybridization makes the identification of a TSS  harder, a TSS bridging the SO-gap is identified close to the $\overline{\Gamma}$ points in InSb and GaAs (top in Fig.~\ref{fig:surf}). These findings support the topological characterization of the SO band given above. In addition to the TSS, a normal surface state (NSS) appears in the calculated surface electronic structures, which is not connected with projected bulk bands.

\section{Experimental aspects}
\label{sec:experiment}
We substantiate our theoretical findings by SX-ARPES experiments on (001)-terminated surfaces of InSb and GaAs. The experiments have been carried out at the I09 beamline of the Diamond Light Source, UK\@. The endstation is equipped with a Specs Phoibos 225 hemispherical electron analyzer. The sample temperature during the measurements was stabilized at about $\unit[10]{K}$. The best total energy resolution was about $\unit[60]{meV}$.

Commercially available (001)-terminated zinc-blende semiconductors InSb and GaAs were cleaned by standard sputter and anneal cycles until a clear  c(8$ \times $2) surface reconstruction was observed by low-energy electron diffraction. After the preparation samples were transferred in UHV suitcase with a base pressure better than $\unit[10^{-10}]{mbar}$ to the beamline for SX-ARPES measurements.

Due to possible strong hybridization of a TSS with bulk states (as predicted by theory) we prefer excitations by soft X-rays rather than by low-energy (vacuum-ultraviolet) photons, as the former provide increased bulk sensitivity  and $k_{\perp}$ resolution \cite{Strocov2003}. Recall that previous SX-ARPES measurements resolved a TSS in the SO-gap of $\alpha$-Sn \cite{Rogalev2017}.

\begin{figure}
	\centering
	\includegraphics[width=0.95\columnwidth]{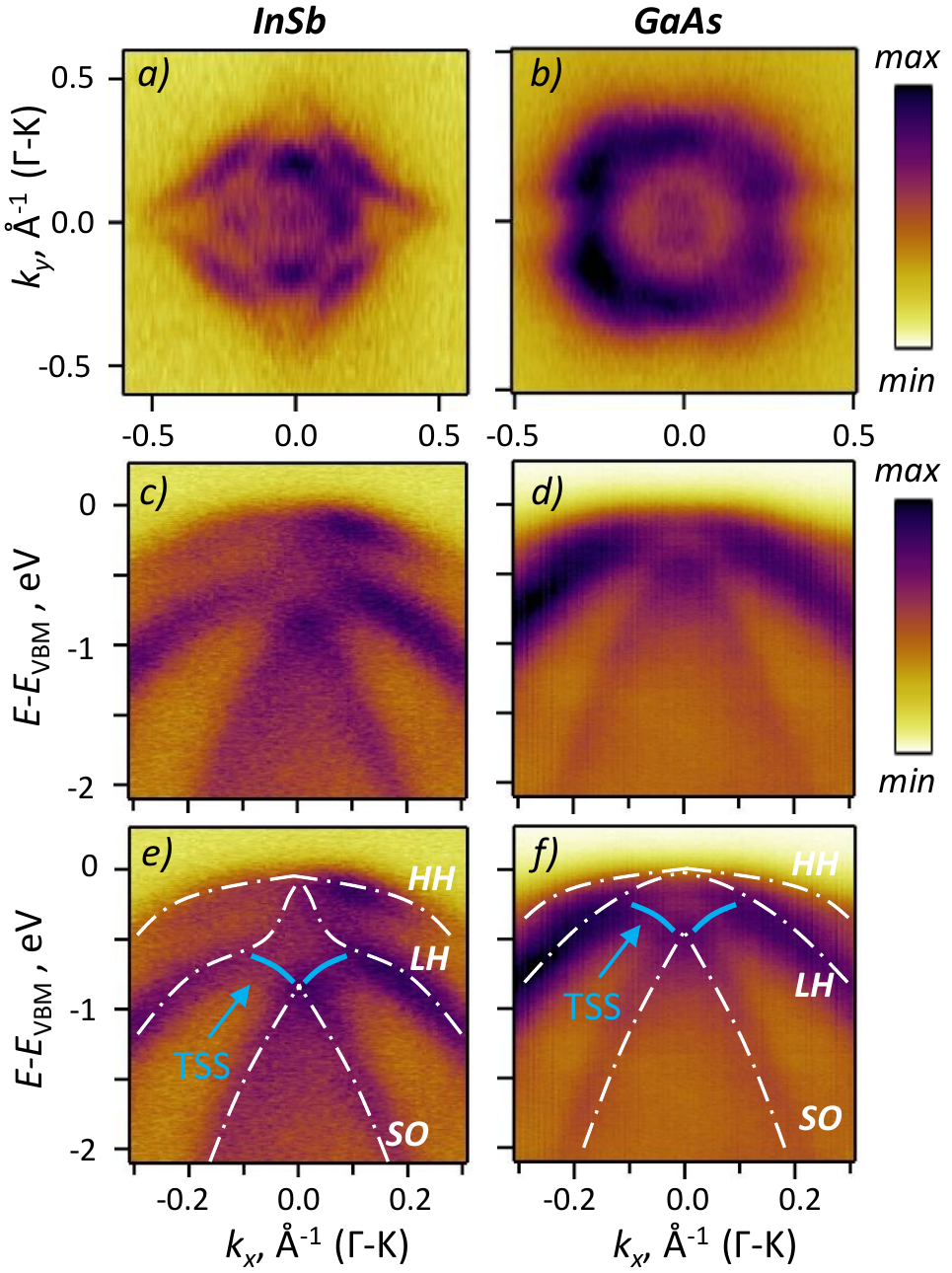}
	\caption{SX-ARPES data measured on (001) surfaces of InSb (left) and GaAs (right) with photon energies $h\nu = \unit[346]{eV}$ (linear p-polarized light) and $h\nu = \unit[164]{eV}$ (circular polarized light), respectively. (a) and (b): constant binding-energy maps at $E - E_{\mathrm{VBM}} = \unit[-0.75 \pm 0.05]{eV}$ for InSb (a) and GaAs (b). (c) and (d): intensity maps $I(E, k_{x}$) along $\Gamma$-- K ($\overline{\Gamma}$--$\overline{\mathrm{X}}$) for InSb (c) and GaAs (d). The spectral weight between the SO and LH bands is ascribed to TSSs which are emphasized by guides for the eye in (e) and (f).}
	\label{fig:SX-ARPES}
\end{figure}

Figure~\ref{fig:SX-ARPES} presents SX-ARPES data measured on (001)-terminated surfaces of InSb and GaAs. The photon energies ($\unit[346]{eV}$ for InSb and $\unit[164]{eV}$ for GaAs) correspond to a $k_{\perp}$ close to the bulk $\Gamma$ points, which was checked experimentally by varying the photon energy. Constant energy maps shown in Figs.~\ref{fig:SX-ARPES}a and~\ref{fig:SX-ARPES}b reveal a fourfold symmetry typical of bulk electronic structure of (001)-terminated InSb and GaAs compounds. Note that p-polarized light used for InSb allows photoelectron excitation only from initial states whose wave function is symmetric (even) with respect to the measurement plane ($(\overline{1}10)$ mirror plane K$\Gamma$Z), since the final state of the photoelectron is even \cite{Hermanson1977}. Thus, only states with dominant $s$, $p_{z}$, and symmetric combinations of $p_{x}$ and $p_{y}$ orbitals appear in Fig.~\ref{fig:SX-ARPES}c. In turn, circular-polarized light used for GaAs (Fig.~\ref{fig:SX-ARPES}d) does not restrict the mirror symmetry of the initial states in the photoemission process.

For InSb, apart from all three well-resolved HH, LH, and SO bulk bands, there is additional distinct spectral weight within the SO-gap between the LH and SO band (Fig.~\ref{fig:SX-ARPES}b, marked as TSS in Fig.~\ref{fig:SX-ARPES}e). The latter agrees very well with the theoretical results presented in Fig.~\ref{fig:surf}, a finding which provides clear experimental confirmation for the existence of a TSS and the non-trivial topology of the SO band.

GaAs also shows traces of an additional state within the SO-gap (Figs.~\ref{fig:SX-ARPES}d and~\ref{fig:SX-ARPES}f), although more pronounced at lower photon energy. This difference can be attributed to the degree of hybridization of the TSS with bulk states in these two materials and the accompanying surface localization of the TSS; a similar effect was observed in $\alpha$-Sn \cite{Rogalev2017}.

\section{Conclusions and outlook}
\label{sec:discussion}
Our theoretical and experimental investigation proves that non-trivial topological bands with a strong TI character exist in the occupied bands of `trivial' diamond and zinc-blende semiconductors whose band structures are standard examples in text books on solid state physics. For the example of InSb, the calculated topological invariants are confirmed according to the bulk-boundary correspondence by identifying the associated TSSs at the (001) surface. These theoretical findings are fully in line with the experimental observations of the TSSs by means of SX-ARPES\@.

In a recent study on the (001) surfaces of Ge and Si by ARPES and \textit{ab-initio} theory \cite{Seo2014} some experimental features remained unexplained. According to the findings presented here, the unresolved bands in that paper ($E_2$ in Fig.~4a) may be attributed to a TSS\@. And considering the similarity of the zinc-blende with the half-Heusler structure, common features in both material classes are likely. Surface states were reported in occupied bands of half-Heusler materials~\cite{Liu2011,Logan2016,Liu2016,Hosen2018} which might be TSSs of a similar non-trivial topological origin as those of InSb and GaAs, which is indeed discussed in Refs.~\onlinecite{Liu2016,Hosen2018}.

A signature of a topological phase transition is the closing and reopening of a band gap upon variation of certain parameters, for example the SOC strength. This mechanism illustrates the non-trivial topology in, say, Bi$_2$Te$_3$ and similar compounds. In the materials considered in this paper, however, the non-trivial topology arises immediately when SOC is introduced in a calculation: the SO-gap does not close and re-open with increasing SOC strength. This scenario can be found for example also in graphene. In Ref.~\onlinecite{Vergniory2018} a vast amount of known materials was classified according to the topological quantum chemistry~\cite{Bradlyn2017}. Among others, two classes of topologically non-trivial materials were found; those with the usual band inversion and those which are either topological insulators (if gapped) or semimetals (if not gapped). The low-lying valence bands of diamond and zinc-blende materials studied in the present work belong to the former.

Since the TSSs in InSb and GaAs cannot be separated from the projected bulk bands, their influence on transport properties is likely marginal. Considering fundamental aspects, however, the present work augments the understanding of non-trivial topology in band structures of materials considered trivial before and extends the common belief of the necessity of a band inversion by the alternative `direct' scenario that we demonstrate for InSb and GaAs.

\begin{acknowledgments}
  This work was supported by the Priority Program SPP 1666 of Deutsche Forschungsgemeinschaft (DFG). We acknowledge financial support from the DFG through the W\"{u}rzburg-Dresden Cluster of Excellence on Complexity and Topology in Quantum Matter -- \textit{ct.qmat} (EXC 2147, project-id 39085490). Diamond Light Source (Didcot, UK) is gratefully acknowledged for beamtime under proposal SI19512.
\end{acknowledgments}

\bibliographystyle{apsrev4-1}
\bibliography{paper.bbl}

\end{document}